\documentclass[twocolumn,showpacs,preprintnumbers,amsmath,amssymb,floatfix]{revtex4}
\usepackage{graphicx}
\usepackage{dcolumn}
\usepackage[latin1]{inputenc}

\begin{document}


\title{Theoretical approach to the ductile fracture of polycrystalline
solids}

\author{Miguel Lagos}
\email{mlagos@utalca.cl}
\affiliation{Departamento de Tecnolog{\'\i}as Industriales, Facultad de
Ingenier\'\i a, Universidad de Talca, Campus Los Niches, Curic\'o,
Chile}
\author{C{\'e}sar Retamal}
\author{Rodrigo Valle}
\affiliation{Departamento de Ingenier{\'i}a y Gesti{\'o}n de la
Construcci{\'o}n, Facultad de Ingenier{\'\i}a, Universidad de Talca,
Campus Los Niches, Camino a los Niches km 1, Curic{\'o}, Chile}

\date{August 26, 2018}

\begin{abstract}
It is shown here that fracture after a brief plastic strain, typically
of a few percents, is a necessary consequence of the polycrystalline
nature of the materials. The polycrystal undergoing plastic deformation
is modeled as a flowing continuum of random deformable polyhedra,
representing the grains, which fill the space without leaving
voids. Adjacent grains slide with a relative velocity proportional to
the local shear stress resolved on the plane of the shared grain
boundary, when greater than a finite threshold. The polyhedral grains
reshape continuously to preserve matter continuity, being the forces
causing grain sliding dominant over those reshaping the grains. It has
been shown in the past that this model does not conserve volume,
causing a monotonic hydrostatic pressure variation with strain. This
effect introduces a novel concept in the theory of plasticity because
determines that any fine grained polycrystalline material will fail
after a finite plastic strain. Here the hydrostatic pressure
dependence on strain is explicitly calculated and shown that has a
logarithmic divergence which determines the strain to fracture.
Comparison of theoretical results with strains to fracture given by
mechanical tests of commercial alloys show very good agreement.
\end{abstract}

\pacs{62.20.F-, 62.20.mm}


\maketitle

\section{Introduction}
\label{introduction}

Asking why things break when subjected to strong enough forces may
sound superfluous because breaking objects is one of the most early
experiences of every person. In reality, explaining why solids
undergoing plastic deformation are unable of achieving a steady flow
regime and collapse past a finite plastic flow, or with almost no flow
at all, is a most important scientific and technical problem yet
unsolved. In technical grounds the point is quite serious because of
the high expenses associated to fatigue and failure of functional
articles. As well, the design of machine parts and structures is
always restricted by the strength of the materials they will be made
of, which puts limits to their efficiency and bounds costs from
below. Since the early investigations of Griffith \cite{Griffith}, who
claimed that the tensile strength of glass is lowered by the presence
of very small pre--existent cracks that concentrate stresses when the
material is loaded, and Irwin
\cite{Irwin} and Orowan \cite{Orowan}, who extended the idea to
ductile solids, a great amount of effort has been expended in
elucidating why solid materials break from the atomic point of view.

Nowadays the question has turned to {\bf how} solids fail, instead of
{\bf why} they break. Certainly, the two issues are closely related
and answering the former question may clarify the latter, but not
necessarily. Most of the contemporary research on this subject relies
on the hypothesis of cracks, and ascribes brittle behavior to the
ability of stressed crack tips to propagate conserving their
atomically sharp edges. In ductile solids the tip of the crack blunts,
broadens and flows, demanding increasing effort to make it progress
\cite{Rice, Thompson, Hirsch, Khanta, Hartmaier, Gumbsch, Green}.
Unfortunately, the problem of stress induced crack propagation has
proven to be exceedingly complex, and neither theory nor computer
simulations \cite{Marder, Henry, Batrouni} have produced conclusive
answers on the fracture process and the origin of brittle or ductile
fracture. The complex evolution of crack growth has been accurately
measured \cite{Sharon1, Sharon2, Fineberg}, confirming atomic scale
model predictions \cite{Ching, Slepyan} that the dynamics of a crack
tip is highly unstable, and steady motion in a given direction is in
most situations impossible.

We show here that the reason {\bf why} continued deformation
inevitably makes solids to break, undergoing either brittle, ductile
or superplastic fracture, is much more basic and simpler than how
fracture proceeds. Resorting to a very general model for the structure
of the solid, we demonstrate in what follows that fine grained
polycrystalline materials are not able of a steady flow, no matter the
strength of the forces involved, and should collapse after reaching a
finite plastic strain.

At a scale much larger than the grain size, polycrystalline matter
lacks symmetry constrictions and periodicity, and displays same
average packing and properties in all directions, and over its whole
extention. Despite this, assimilating an even very fine grained
polycrystal to an homogeneous and isotropic continuum may lead to
gross errors, no matter the scale, when dealing with it as a dynamical
medium. The faceted nature of the structural constituents of a
polycrystal determines that the force fields governing their plastic
flow yield $\nabla\cdot\vec v\ne 0$, where $\vec v$ is the velocity
field of the material continuum. This means that flow makes the
specific volume to vary. Grain elasticity in polycrystalline solids
allows for some density variation, and hence the medium can flow up to
some limit, yielding ductile behaviour. However, the consequent
pressure build up influences strongly the ongoing deformation, which
cannot be steady, and finally produces fracture. Thus ductility is
closely related to compressibility.

\section{Theory}\label{theory}
\subsection{The force model}\label{model}

The model for the plastic flow of a polycrystalline solid has been
extensively studied, principally in the context of superplasticity,
but is expected to equally hold for normal ductile solids. However, a
brief account of its physical basis and the resulting general
theoretical scheme is in order here.

The plastic deformation of a fine grained polycrystalline solid is
modelled as a flowing continuum of random irregular polyhedra of
different shapes and sizes, representing grains, which share faces.
The model is essentially the same as the one of Ref.~\cite{Lagos0}.
Grains can move over long paths by sliding along the shared surfaces,
or grain boundaries, accommodating effortlessly their shapes to
preserve matter continuity. Certainly, grain shape accommodation
demands some effort, but it is assumed much smaller than the one
required for grain sliding. In other words, the shear stress between
two sliding grains is greater than the critical resolved shear stress
(CRSS) demanded by slip deformation of the crystallites. This way,
grain boundary sliding is the rate limiting process in the plastic
strain. In the present scheme grains always retain their individuality
and mass, and are the dynamical entities.  The flow is driven by a
field of tensor forces between the grains, determined by the stress
tensor.

\begin{figure}[h!]
\begin{center}
\includegraphics[width=6cm]{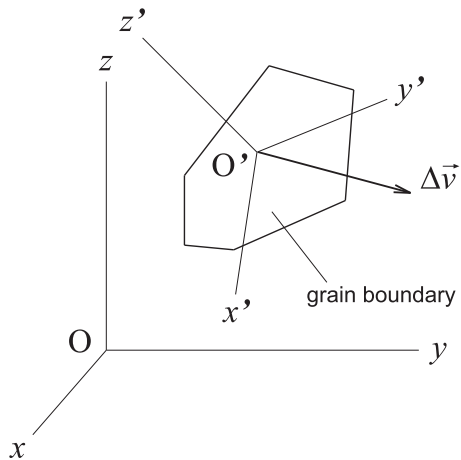}
\caption{\label{Fig1} Local reference system $(x^\prime y^\prime
z^\prime)$, with the $z^\prime$ axis normal to the plane of the common
boundary of two adjacent grains. The relative velocity $\Delta\vec v$
of the two grains is in the $x'y'$ plane. The axes of the $(xyz)$
frame of reference are in the principal directions of the stress
tensor.}
\end{center}
\end{figure}

Fig.~\ref{Fig1} shows a local frame of reference $(x'y'z')$ with the
$x'y'$ plane coincident with the boundary between two adjacent grains.
The total shear stress in the shared boundary plane then reads
$\tau_{z'}=(\sigma_{x'z'}^2+\sigma_{y'z'}^2)^{1/2}$, where
$\sigma_{i'j'}$, $i',j'=x',y',z'$, stands for the components of the
stress tensor in this local coordinate system. There is strong
evidence that the sliding relative speed $|\Delta\vec v|$ of two
adjacent grains obeys a linear law of the general form $|\Delta\vec v|
=\mathcal{Q}(\tau_{z'}-\tau_c)$ for $\tau_{z'}>\tau_c$ in plastic
deformation \cite{Lagos0,Qi,Fukutomi,Bellon,Lagos1,Lagos2}. Here
$\mathcal{Q}$ is a proportionality coefficient and $\tau_c$ is a
critical shear stress such that $|\Delta\vec v|=0$ when
$\tau_{z'}\le\tau_c$. As $\Delta\vec v$ is parallel to the shear force
in the plane of the interface, its components are given by $\Delta
v_{i'}=\mathcal{Q}(\tau_{z'}-\tau_c)(\sigma_{i'z'}/\tau_{z'})$,
$i'=x',y'$, for $\tau_{z'}\ge\tau_c$. This expresion for $\Delta\vec
v$ has proven to hold with great accuracy for several aluminium,
titanium and magnesium alloys
\cite{Lagos2,LagosRetamal1,LagosRetamal2}. Hence the force law at the
grain scale reads

\begin{equation}
\begin{aligned}
&\Delta v_{i'}=
\begin{cases}
\mathcal{Q}\,
\left( 1-\dfrac{\tau_c}{\tau_{z'}}\right) \sigma_{i'z'},\,
&i'=x', y', \, \text{if}\, \tau_{z'}>\tau_c \\
0, &\text{ otherwise},  
\end{cases} \\
&\Delta v_{z'}\equiv 0. \\
\label{E1}
\end{aligned}
\end{equation}

\noindent
The coefficient $\mathcal{Q}=\mathcal{Q}(p,T)$ does not depend on the
shear stresses and neither on the orientation of the grain boundary,
therefore its dependence on the normal stresses is only via the
hydrostatic pressure invariant 

\begin{equation}
p=-(\sigma_{x'x'}+\sigma_{y'y'}+\sigma_{z'z'})/3 .
\label{E2}
\end{equation}

The next step is to express the force law (\ref{E1}) in the frame of
reference $(xyz)$, common to all grain surfaces, instead of the local
ones $(x^\prime y^\prime z^\prime)$. Given the rotation matrix
$R(\theta ,\phi)=(R_{ij}(\theta ,\phi))$ connecting the two frames one
can put the local stress tensor

\begin{equation}
(\sigma_{i^\prime j^\prime})
=R(\theta ,\phi)(\sigma_{ij})R^T(\theta ,\phi)
\label{E3}
\end{equation}

\noindent
in terms of the stress tensor $(\sigma_{i j})$ of the externally
applied forces and the Euler angles $(\theta ,\phi)$ of the grain
boundary plane. The macroscopic force law is obtained from replacing
in Eq.~(\ref{E1}) and averaging over the Euler angles. Invoking also
Hooke's law one has that

\begin{equation}
\nabla\cdot\vec v=\frac{\dot V}{V}=-\frac{\dot p}{B},
\label{E4}
\end{equation}

\noindent
where $B$ is the bulk elastic modulus, $\dot p$ the pressure variation
rate, and $\dot{V}/V$ the volume variation rate per unit volume. A
detailed account of the procedure would be in excess here because can
be found in the literature
\cite{Lagos0,Lagos1,Lagos2,LagosRetamal1,LagosRetamal2}.

\subsection{The equations of motion}\label{motion}

After a rather tedious set of mathematical steps
\cite{Lagos0,Lagos2,LagosRetamal1,LagosRetamal2,LagosConte1} the
procedure outlined above for the special case of an externally applied
unidirectional normal stress $\sigma$ on a polycrystalline solid,
isotropic in the scale much larger than the mean grain size $d$,
yields the complete set of macroscopic equations of motion

\begin{equation}
\dot\varepsilon =
s\,\frac{\tau_c\mathcal{Q}(p)}{2d}\left[ \cot (2\theta_c)
+2\theta_c-\frac{\pi}{2}\right] ,
\label{E5}
\end{equation}

\begin{equation}
\begin{aligned}
\dot p= &sB\frac{\tau_c\mathcal{Q}(p)}{2d}
\bigg[ \frac{1 -\cos(2\theta_c)}{\sin (2\theta_c)}
-2\theta_c\bigg(1+\frac{2}{\pi \sin (2\theta_c)}\bigg)\\
&-\frac{2}{\pi}\cos (2\theta_c)+\frac{\pi}{2}\bigg],
\label{E6}
\end{aligned}
\end{equation}

\noindent
where $\dot\varepsilon$ is the strain rate in the direction of the
applied stress $\sigma$, $s=\pm 1$ assumes the positive and negative
values for tension and compression, respectively, and the auxiliary
variable $\theta_c$ is given by

\begin{equation}
\sin (2\theta_c)=\frac{4\tau_c}{3|\sigma +p|}.
\label{E7}
\end{equation}

The properties of the specific material enters the theoretical
formulation through the coefficient $\mathcal{Q}(p,T)$, governing
grain boundary sliding. It has been studied in detail for fine grained
polycrystalline solids and has been shown to be of the general form

\begin{equation}
\frac{\mathcal{Q}(p,T)}{4d}=C_0\frac{\Omega^*}{k_BT}
\exp\left( -\frac{\epsilon_0+\Omega^*p}{k_BT}\right)
\label{E8},
\end{equation}
 
\noindent
where $k_B$ is the Boltzmann constant, $T$ the absolute temperature,
the coefficient $C_0$ depends only on the grain size $d$, the constant
$\epsilon_0$ is the energy necessary for evaporating a crystal vacancy
from the grain boundary, and $\Omega^*$ is the excitation volume for
the same process.

Eqs.~(\ref{E5}), (\ref{E6}) and (\ref{E7}) show that plastic flow is
essentially a time dependent problem. They govern the coupled time
evolution of the three variables, $\sigma$, $\varepsilon$ and $p$,
relevant for the cylindrically symmetric deformation of a
polycrystalline continuous medium. The actual behaviour of these
variables in specific circumstances depends also on the initial
conditions and deformation path ($\sigma =\text{constant}$,
$\dot\varepsilon =\text{constant}$, or any other imposed condition
between the variables and their time derivatives). The observed
dependence on history of the plastic properties of ductile solids is
usually attributed to structural variations or deformation induced
damage. In the present scheme, history enters through the initial
condition for the variable $p$, which is omitted in the traditional
theoretical approaches to plasticity. Here, the system described by
Eqs.~(\ref{E5}), (\ref{E6}) and (\ref{E7}) behaves always the same
way, but its evolution depends on the initial conditions for the
variables, which include $p$ \cite{LagosConte1}. In opposition to the
classical theory of plasticity, it exists a nontrivial transversal
stress $\sigma_\perp=-(\sigma +3p)/2$ which is not an independent
variable, but evolves in time as dictated by the equations of motion.
One can set $\sigma_\perp =0$ as a natural initial condition if the
material has been previously annealed, but $\sigma_\perp$ is expected
to take finite values on the subsequent deformation. As the magnitude
inside the square brackets in the right hand side of Eq.~(\ref{E6}) is
positive for any $\theta_c$, $\dot p$ has the sign of $s$. The
transversal stress $\sigma_\perp$ then decreases monotonically to
negative values for positive $\sigma +p$. Physically, this means that
the plastic stretching in one direction is always accompanied by a
finite compression in the plane normal to the deformation axis, which
increases monotonically with strain. This explains why necking always
precedes ductile fracture \cite{LagosConte1}.

\subsection{The equations for constant strain rate}
\label{constant}

Replacing Eq.~(\ref{E6}) in the identity $d\varepsilon
=(\dot\varepsilon /\dot p)\, dp$ one has that

\begin{equation}
\begin{aligned}
d\varepsilon =&s\frac{2\dot\varepsilon d}{B\tau_c}
\bigg[\frac{1-\cos (2\theta_c)}{\sin (2\theta_c)}
-2\theta_c\bigg(1+\frac{2}{\pi\sin (2\theta_c)}\bigg)\\
&-\frac{2}{\pi}\cos (2\theta_c)
+\frac{\pi}{2}\bigg]^{-1}\frac{dp}{\mathcal{Q}(p,T)},
\label{E9}
\end{aligned}
\end{equation}

\noindent
where $\dot\varepsilon$ is considered as a given constant. As long as
$\dot\varepsilon =\text{constant}$, Eq.~(\ref{E5}) shows that
$\theta_c=\theta_c(p)$. Combining the derivatives of Eqs.~(\ref{E5})
and (\ref{E8}) with respect to $p$ it can be shown that

\begin{equation}
\frac{dp}{\mathcal{Q}}=
-s\frac{\tau_ck_BT}{\Omega^*\dot\varepsilon d}
\cot^2 (2\theta_c)\, d\theta_c .
\label{E10}
\end{equation}

\noindent
Replacing now Eq.~(\ref{E10}) in (\ref{E9}) and integrating, it is
finally obtained

\begin{equation}
\begin{aligned}
\varepsilon =
&-\frac{2k_BT}{B\Omega^*}\int_{\theta_0}^{\theta_c}d\theta\,
\cot^2(2\theta_c)\bigg[\frac{1-\cos (2\theta)}{\sin (2\theta)}\\
&-2\theta\bigg(1+\frac{2}{\pi\sin (2\theta)}\bigg)
-\frac{2}{\pi}\cos (2\theta )+\frac{\pi}{2}\bigg]^{-1},
\label{E11}
\end{aligned}
\end{equation}

\noindent
where the limits $\theta_0$ and $\theta_c$ correspond to the critical
angles for the initial and final values of the strain, $\varepsilon =0$
and $\varepsilon$, respectively. This way, $\varepsilon$ is related
with the auxiliary variable $\theta_c$ by an expression of the form

\begin{equation}
\varepsilon =\frac{k_BT}{B\Omega^*}[F(\theta_c)-F(\theta_0)]
\quad (\dot\varepsilon =\text{ constant}),
\label{E12}
\end{equation}

\noindent
where $F(\theta)$ is the universal function

\begin{equation}
\begin{aligned}
F(\theta)&=-2\int_{\pi /8}^{\theta} d\theta\,
\cot^2(2\theta_c)\bigg[\frac{1-\cos (2\theta)}{\sin (2\theta)}\\
&-2\theta\bigg(1+\frac{2}{\pi\sin (2\theta)}\bigg)
-\frac{2}{\pi}\cos (2\theta )+\frac{\pi}{2}\bigg]^{-1},
\label{E13}
\end{aligned}
\end{equation}

\noindent
which is monotonically decreasing in its whole range $(0,\pi/4)$ and
has two singularities, at $\theta =0$ and $\theta =\pi/4$. If the
material has been thoroughly annealed prior to the plastic deformation,
it holds the initial condition $p=-\sigma_0/3$ at $\varepsilon =0$,
where $\sigma_0$ is the stress at the beginning of the plastic
deformation.

The magnitude of $\varepsilon$ is controlled by the adimensional
coefficient appearing in Eqs.~(\ref{E11}) and (\ref{E12}), which is a
very small quantity. The bulk modulus $B$ for metals is of the order
of $10^{11}\,\text{Pa}$. Previous literature on aluminium and titanium
alloys shows that $\Omega^*$ is $2.6\times 10^{-27}\,\text{m}^3$ for
Al--8090 and $5.9\times 10^{-28}\,\text{m}^3$ for titanium Ti--6Al--4V
at rather high temperatures \cite{Lagos2}.  Assuming $\Omega^*$ does
not vary too much with $T$ one can take these figures to estimate that,
at $T=300\,\text{K}$,

\begin{equation}
\frac{k_BT}{B\Omega^*}\thicksim
2.3\times 10^{-5}\, -\, 7.0\times 10^{-5}.
\label{E14}
\end{equation}

\noindent
Because of the small value of the coefficient (\ref{E14}), any
significant strain $\varepsilon$ demands that the function $F(\theta)$
be large, of the order of $10^3$ to have a strain of a few percents.
Hence $\theta$, or $\theta_c$, or both, must be in one of the two
asymptotic regions $\theta\gtrsim 0$ or $\theta\lesssim\pi/4$. The
threshold stress $\tau_c$ for grain sliding is generally in the range
$0.5-5\, \text{MPa}$, i.~e.~much smaller than the applied stresses
$\sigma$ that are customary in mechanical tests. Hence the divergence
at $\theta =0$ should be the right one and appreciable strains occur
for

\begin{equation}
\theta_c(\varepsilon ,\dot\varepsilon ,T)\approx 0.
\label{E15}
\end{equation}

\noindent
The other pole of function $F(\theta)$ corresponds to very slow
flux, as occurring in superplastic deformation.

\subsection{Theory in the first order in $\theta_c$}\label{first}

Up to the first order in $\theta$ the expression in between the square
brackets in Eqs.~(\ref{E11}) and (\ref{E13}) reduces to

\begin{equation}
\begin{aligned}
\bigg[\frac{1-\cos (2\theta )}{\sin (2\theta )}
-2\theta\bigg( 1&+\frac{2}{\pi\sin (2\theta )}\bigg) \\
-\frac{2\cos (2\theta )}{\pi}&+\frac{\pi}{2}\bigg]\approx
\frac{\pi}{2}-\frac{4}{\pi}-\theta .
\end{aligned}
\label{E16}
\end{equation} 

\noindent
The constant $\pi /2-4/\pi =0.29756$ is not small enough and we can
neglect $\theta$ when compared with it. Thus, with no significant
lost of precision the exact equation 

\begin{equation}
\begin{aligned}
&\frac{dp}{d\varepsilon}=
sB\frac{\tau_c\mathcal{Q}(p,T)}{2\dot\varepsilon d}
\bigg[ \frac{1 -\cos(2\theta_c)}{\sin (2\theta_c)}\\
&-2\theta_c\bigg(1+\frac{2}{\pi \sin (2\theta_c)}\bigg)
-\frac{2}{\pi}\cos (2\theta_c)+\frac{\pi}{2}\bigg],
\label{E17}
\end{aligned}
\end{equation}

\noindent
can be reduced to the much simpler first order differential equation

\begin{equation}
\frac{dp}{d\varepsilon}=s\left(\pi-\frac{8}{\pi}\right)
\frac{C_0B\tau_c\Omega^*}{k_BT\dot\varepsilon}
\exp\left( -\frac{\epsilon_0+\Omega^*p}{k_BT}\right),
\label{E18}
\end{equation}

\noindent
whose solution can be written as 

\begin{equation}
\begin{aligned}
p-p_0=\frac{k_BT}{\Omega^*}\ln &\bigg[ 1-
C_0\frac{\pi^2-8}{\pi}\frac{\tau_c B}{\dot\varepsilon}
\bigg(\frac{\Omega^*}{k_B T}\bigg)^2 \\
&\times\exp\bigg( -\frac{\epsilon_0+\Omega^*p_0}{k_B T}\bigg)
|\varepsilon|\bigg]\, .
\label{E19}
\end{aligned}
\end{equation}

\noindent
where it was substituted $s\varepsilon =|\varepsilon|$.

Eq.~(\ref{E19}) expresses the main finding of this work: when the
modulus $|\varepsilon|$ of the strain approaches from below the value

\begin{equation}
\varepsilon_{\text{frac}}=
\frac{\pi\dot\varepsilon}{(\pi^2-8)C_0\tau_c B}
\bigg(\frac{k_B T}{\Omega^*}\bigg)^2
\exp\bigg( \frac{\epsilon_0+\Omega^*p_0}{k_B T}\bigg)
\label{E20}
\end{equation}

\noindent
the hydrostatic pressure $p$ diverges logarithmically. According to
the definition (\ref{E2}) positive stresses (tension) contribute
negatively to the hydrostatic pressure $p$. If the sample is
conveniently annealed prior to the tensile test then
$p_0=-\sigma_0 /3$, where $\sigma_0$ is the applied initial tensile
stress. As the test proceeds, $p=-(\sigma +2\sigma_\perp )/3$ increases
monotonically with $\varepsilon$, and the transversal stress
$\sigma_\perp$ increases from zero to negative (compressive) values.
When $\varepsilon$ approaches the critical value
$\varepsilon_{\text{frac}}$ the transversal stress $\sigma_\perp$
increases very rapidly, producing the characteristic neck and fracture.
Therefore, Eq.~(\ref{E20}) for $\varepsilon_{\text{frac}}$ expresses
the strain to fracture of the material.
 
\subsection{Necking and strain to fracture}\label{frac}

Eq.~(\ref{E20}) gives the strain to fracture in terms of the constants
of the theory. However one can express it in terms of more standard
coefficients and easily measurable quantities. Combining
Eqs.~(\ref{E5}), (\ref{E8}), and taking into account the
asymptotic approximation (\ref{E15}) to write

\begin{equation}
\cot (2\theta_0) +2\theta_0 -\frac{\pi}{2}\approx \frac{1}{2\theta_0}
\approx\frac{\sigma_0}{2\tau_c}\, ,
\label{E21}
\end{equation}

\noindent
Eq.~(\ref{E20}) can be written as

\begin{equation}
\varepsilon_{\text{frac}}=
\frac{\pi}{(\pi^2-8)}\frac{k_B T}{B\Omega^*}
\frac{\sigma_0}{\tau_c}.
\noindent
\label{E22}
\end{equation}

\noindent
We recall that $\sigma_0$ is the stress registered when the plastic
deformation at the chosen constant strain rate $\dot\varepsilon$
begins. The bulk modulus $B$ is in tables and the only undetermined
parameter is the product $\Omega^*\tau_c$. However, $\Omega^*\tau_c$
can be determined independently from other features of the plastic
deformation of the sample in order to have a parameter free test of
Eq.~(\ref{E22}). To show how well this expression compares with
experiment, we include next a study of a representative commercial
steel.

\begin{figure}[h!]
\begin{center}
\includegraphics[width=8cm]{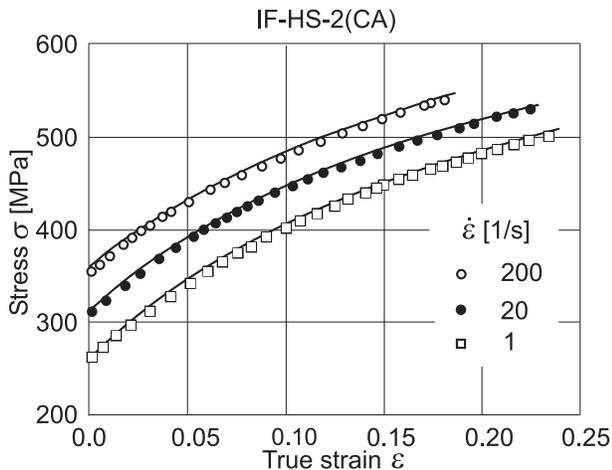}
\caption{\label{Fig2} Circles represent the stress--strain experimental
data for a copper--alloyed high--strength interstitial free steel at
the three strain rates shown in the inset \cite{Rana}. The continuous
lines represent the predictions of Eq.~(\ref{E11}) with the parameters
optimizing the fit to the experimental points, shown in Table
\ref{table1}.}
\end{center}
\end{figure}

Figure \ref{Fig2} shows the results of a mechanical test of a
copper--alloyed high--strength interstitial free steel at strain rates
$1$, $20$ and $200\,\text{s}^{-1}$ \cite{Rana}, together with the fits
of Eq.~(\ref{E11}) with the asymptotic approximation (\ref{E15}). The
high quality of the agreement between theory and experiment is apparent
in the figure, and the very little dispersion of the fitting parameters
$\Omega^* B$ and $\tau_c$ shown in Table \ref{table1} reinforces this
perception. The last column of Table \ref{table1} displays the strain
to failure $\varepsilon_{\text{frac}}$ for the three strain rates, as
given by Eq.~(\ref{E22}) where the parameters appearing in the left
side of Table \ref{table1} were substituted. The values are very close
to those measured in the mechanical testings. Comparisons between
predicted strains to fracture with published results of experimental
tests for many other commercial alloys exhibit same agreement as the
one shown in Fig.~\ref{Fig1} and Table \ref{table1}.

\begin{center}
\begin{table}[h!]
\caption{\label{table1} Values for the parameters giving the fits
of Fig.~\ref{Fig2} and calculated strains to failure.}
\vskip 12pt
\begin{center}
\begin{tabular}{|cccc|c|}
\hline
$\dot\varepsilon\,[\text{s}^{-1}]$ & $\dfrac{k_BT}{\Omega^*B}$ &
$\tau_c\,[\text{MPa}]$ & $\dfrac{\tau_c}{\sigma_0}$ &
$\varepsilon_{\text{frac}}$ \\
\hline
200 & $8.03\times 10^{-2}$ & 222 & 0.618 & 0.218 \\
20 & $7.43\times 10^{-2}$ & 208 & 0.667 & 0.187 \\
1 & $1.06\times 10^{-1}$ & 219 & 0.846 & 0.211 \\
\hline
\end{tabular}
\end{center}
\end{table}
\end{center}

\section{Conclusions}

Although the existence of cracks and imperfections inside a stressed
solid may contribute to accelerate fracture, the general cause of
ductile fracture is not in them. An ideal fine--grained polycrystalline
material, free of voids and cracks, whose grains are prone to slide,
readily accommodating each other's shapes, inevitably should fail
after a finite plastic strain. The reason is an elementary condition
that was advanced some years ago \cite{Lagos0} but omitted in other
studies: whatever the mechanisms for stress--dependent grain boundary
sliding and grain shape accommodation may be, they must be consistent
with density conservation to produce a steady flow. However, it is
shown here that if the local shear stresses resolved in the planes of
grain interfases have a finite threshold for causing grain sliding,
density is not conserved in the overall plastic flow. The grains are
increasingly compressed as the sample is being stretched, and hence
grain sliding can only proceed at the expenses of elastic volume
variations of the crystallites.  Fracture after a brief plastic strain,
typically of a few percents, is a necessary consequence of the
polycrystalline nature of the materials. The model gives a simple and
precise closed--form equation for the strain to fracture, which is the
strain at which the internal hydrostatic pressure diverges.

\end{document}